**Droplet Removal by Capillary Lifting**

Ke Sun,[1] Jack R. Panter,[2] Alvin C. M. Shek,[1] Yonas Gizaw,[3] Kislon Voïtchovsky,[1*] and Halim Kusumaatmaja[4†]

ORCID:
Ke Sun: 0000-0003-4144-7485
Jack. R. Panter: 0000-0001-8523-7629
Alvin C. M. Shek: 0000-0002-9222-4065
Yonas Gizaw: 0009-0009-1526-1303
Kislon Voïtchovsky: 0000-0001-7760-4732
Halim Kusumaatmaja: 0000-0002-3392-9479

[1]Department of Physics, Durham University, Durham, DH1 3LE, United Kingdom
[2]School of Engineering, Mathematics and Physics, University of East Anglia, Norwich, NR4 7TJ, United Kingdom
[3]GreenChemSolution, West Chester, OH 45069, United States of America
[4]Institute for Multiscale Thermofluids, School of Engineering, The University of Edinburgh, Edinburgh, EH9 3FB, United Kingdom

The removal of liquid droplets from solid surfaces is central to cleaning, coatings and oil recovery. Here we investigate liquid droplets capillary lifted by an immiscible working liquid. The rising working liquid triggers the formation of a capillary bridge between the solid and the air interface, which can lead to full, partial, or no droplet dewetting. Our theoretical model predicts, and experiments confirm, that the effectiveness of droplet removal can be tuned by manipulating the droplet contact angle with the solid and the interfacial tensions at play. Significantly, dewetting can be enhanced by employing working liquids with high interfacial tension, in contrast to common surface cleaning strategies where surfactants are used to reduce interfacial tension. Our findings can open new avenues for droplet manipulation with reduced resources and more sustainable environmental impact.

Droplet dewetting refers to the process where a liquid droplet retracts its contact from a solid surface. The phenomenon is central to a wide range of engineering and industrial applications, including self-cleaning surfaces [1], inkjet printing [2], and surface patterning [3,4]. To date, most studies of droplet dewetting focus on binary fluid systems whereby the droplet is surrounded by air or an immiscible liquid. In such cases, a large body of literature has shown that liquid retraction can be driven by several mechanisms, such as the droplet interfacial energy in the case of droplet impact [5], spinodal dewetting [6], loss of volume in the case of evaporation [7] or change of effective contact angle due to external fields [8,9].

In contrast, significant gaps remain in our understanding of dewetting in ternary systems, which involves three fluid components and a solid. Yet ternary dewetting is arguably more common than its binary counterpart, observed when a working liquid is used to induce droplet retraction, including for advanced oil recovery (immiscible water-alternating-gas displacement process) [10], surface coatings [4], hair dyeing [11] and various forms of surface cleaning [1,12–14]. Moreover, in surface cleaning, there are strong economic and environmental pressures to devise new ways to displace soils (liquids or other forms of contaminants) from surfaces using less water, energy and surfactants [15–17]. This is further compounded by the emerging global water scarcity [18] and the toxicity and low degradability of synthetic surfactants [19,20]

In this study, we explore a fundamental mechanism of ternary dewetting that we term capillary lifting [Fig. 1(a)]. Here, a 'soil' droplet surrounded by air is lifted from the surface of a solid by the working liquid. The droplet can be simple, composed of pure chemicals (e.g. alkanes); or chemically heterogeneous, as often seen in real-world contexts (e.g. olive oil). Using a bespoke fluidic device (see Fig. S1 of the Supplemental Material SM (1)), the droplet (~10 μL) experiences the slow rise of the working liquid from the surface of the solid (experimental details in SM (1)). The rise velocity is typically ~5 × 10$^{-7}$ m/s. At such small velocity, the system follows a quasi-equilibrium trajectory. As the working liquid contacts the droplet, the latter can be lifted due to capillary effects and fully or partially dewet from the surface [Fig. 1(b)]. Alternatively, it can remain bound to the surface [Fig. 1(c)].

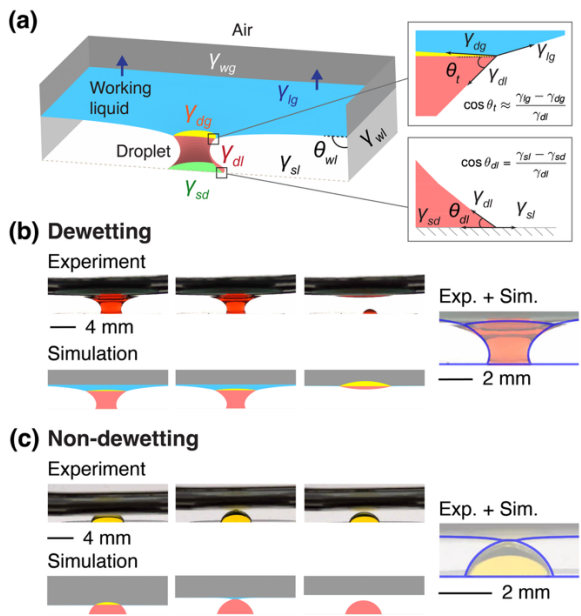

FIG 1. (a) Schematic diagram of a soil droplet being lifted from a solid substrate by a rising working liquid. The relevant thermodynamics parameters are the interfacial tensions (or energies) of the solid substrate-droplet $\gamma_{sd}$ (green), droplet-gas $\gamma_{dg}$ (yellow), liquid-gas $\gamma_{lg}$ (blue), substrate-liquid $\gamma_{sl}$ (white), droplet-liquid $\gamma_{dl}$ (red), liquid-wall $\gamma_{lw}$ (white), and wall-gas $\gamma_{wg}$ (grey). The receding contact angle of the droplet in liquid is $\theta_{dl}$ and the effective top contact angle is $\theta_t$, which can be expressed in terms of the balance of interfacial tensions (see insets). (b-c) Examples of capillary lifting observed experimentally and in simulations. (b) Dewetting of a dyed tetradecane droplet on a polymethyl methacrylate (PMMA) surface by water. (c) Dewetting of a dyed water droplet on PMMA by tetradecane. In (b-c), combined simulation and experimental results are shown with the simulated profiles as blue lines atop the experimental results.

*Contact author: kislon.voitchovsky@durham.ac.uk

†Contact author: halim.kusumaatmaja@ed.ac.uk

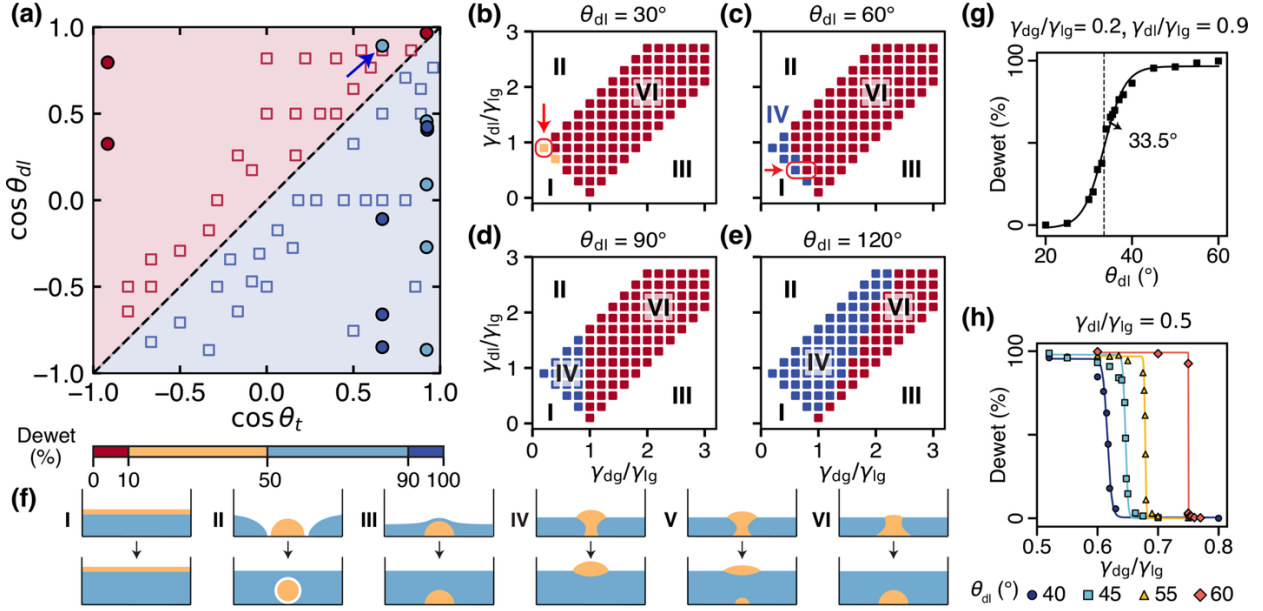

FIG 2. (a) Ternary dewetting simulations (□) and experiments (●) shown as a function of the droplet apparent contact angle at the air interface ($\theta_t$) and solid surface ($\theta_{dl}$). The background color indicates the theoretical prediction for dewetting ($\cos \theta_t > \cos \theta_{dl}$) and non-dewetting ($\cos \theta_t < \cos \theta_{dl}$). The color for the experimental data points indicates the level of dewetting achieved. (b-e) Simulations detailing the dewetting phase diagrams across a range of $\theta_{dl}$ and interfacial tensions $\gamma_{dg}$ (droplet-gas), $\gamma_{dl}$ (droplet-liquid), and $\gamma_{lg}$ (liquid-gas). Schematics of the six regimes are represented in (f), with the droplets in yellow, the working fluid in blue, and the air in white. (g) Dewetting behavior for an interfacial system ($\gamma_{dg}/\gamma_{lg} = 0.2$, $\gamma_{dl}/\gamma_{lg} = 0.9$, red arrow in (b)) as a function of $\theta_{dl}$. (h) Transition boundary from dewetting to non-dewetting at $\gamma_{dl}/\gamma_{lg} = 0.5$ with varying $\gamma_{dg}/\gamma_{lg}$. Results for different $\theta_{dl}$ are shown, including for $\theta_{dl} = 60°$ (red arrow in (c)).

From Fig. 1(b-c), we observe that water can remove >90% of a tetradecane droplet, but tetradecane cannot displace a water droplet. This observation may seem surprising since tetradecane exhibits near-complete wetting on PMMA in air (contact angle ∼ 0°) while water droplets show a significantly higher contact angle of 71 ± 4° [SM (2)]. Tetradecane therefore has the strongest affinity for PMMA. The key to understand this apparent contradiction is that effective capillary lifting requires a high interfacial tension for the working liquid. Here, water has a significantly higher interfacial tension (72 mN/m) than tetradecane (26.56 mN/m). Thus, for such ternary dewetting, the use of surfactants and other small molecules to reduce interfacial tension, ubiquitous in conventional surface cleaning strategies, can actually degrade performance. The fact that capillary lifting mechanism benefits from the absence of surfactants opens an effective avenue for an alternative droplet removal and manipulation approach that is more sustainable and environmentally friendly.

This finding raises several questions. First, what are the key parameters controlling the lifting mechanism and its efficiency? Second, is the transition between a dewetting [Fig. 1(b)] and non-dewetting situation [Fig. 1(c)] continuous when varying the balance of interfacial energies between the media. Third, can the capillary lifting mechanism be extended to real-world complex soils that are chemically heterogeneous?

To better understand the capillary lifting mechanism, we complement our experiments with computer modelling using Surface Evolver [21]. Since the capillary length is comparable to the soil droplet size, gravitational force is included in our analysis [SM (3)] together with the relevant interfacial tensions. The total energy of the system can be expressed as:

$$E_{Total} = \sum_{i,j} \gamma_{ij} A_{ij} + g \sum_k \int_{V_k} \rho_k z dV \quad (1)$$

where $\gamma_{ij}$ represents interfacial tensions between two fluids $i$ and $j$, $A_{ij}$ is the corresponding interfacial area, $g$ is the gravitational acceleration, $z$ is the vertical coordinate, $\rho_k$ and $V_k$ denote densities and volumes of the fluids involved in the system. For simplicity, we take a quasi-static approximation, corresponding to the limit of $Ca \to 0$ (typically, $Ca \sim 10^{-8}$ in our experiments and we do not observe significant dynamic effects for $Ca < 10^{-3}$). In this limit, we slowly and incrementally increase the volume of the invading working liquid and carry out energy minimization for each corresponding volume, see SM (4) for additional details of the numerical scheme. The simulated evolution of the interface configurations agree well with experiments, as shown in Fig. 1(b-c).

*Contact author: kislon.voitchovsky@durham.ac.uk

†Contact author: halim.kusumaatmaja@ed.ac.uk

Fig. 2 summarizes the dewetting outcome for our simulations and experiments for wide-ranging systems. Experimentally, investigated soil droplets range from pure liquids such as tetradecane and water to real-world contaminants including squalane (used in cosmetics and base oil lubricant), olive oil, and used engine oil. For the working fluid, we used either pure liquids or low-concentration glycol ether formulation relevant to industrial applications (0.45 wt% tripropylene glycol methyl ether, 0.45 wt% diethylene glycol hexyl ether and 0.45 wt% ethylene glycol hexyl ether in water). We tested capillary lifting on commercial substrates with varying surface energies, including polytetrafluoroethylene (PTFE), polycarbonate, PMMA, glass, and stainless steel. The liquid interfacial tensions and static contact angles were experimentally determined using pendent and sessile drop setups, as detailed in SM (5-6).

From Fig. 2(a), it is clear that the dewetting and non-dewetting regions follow a trend that can be described theoretically as a function of $\theta_{dl}$ and $\theta_t$ (see Fig. 1a). This can be rationalized from simple geometrical considerations by analogy with capillary bridges. A bridge stability is determined by the contact angles on the top and bottom substrates, here by analogy $\theta_{dl}$ and $\theta_t$. Dewetting is favored when $\cos\theta_{dl} < \cos\theta_t$ (bottom right corner in Fig. 2(a)), as the bottom contact radius vanishes faster than the top contact radius [22,23]. This simple approximation yields an excellent agreement between experiments, simulations and theory, though small deviations are observed close to the boundary line since the air interfaces are slightly curved [SM (7)]. A clear outlier (blue arrow in Fig. 2(a)) occurred when using a formulated solution used as the working fluid instead of a pure liquid. This can be accounted by considering molecular migration and will be discussed subsequently in Fig. 3.

All possible morphologies from the capillary lifting mechanism are shown in Fig. 2(b-e), which capture regime diagrams in the space of interfacial tension ratios $\gamma_{dl}/\gamma_{lg}$ and $\gamma_{dg}/\gamma_{lg}$ for four different values of $\theta_{dl}$. For each diagram, six regions can be broadly defined, labelled I to VI and illustrated as cartoon in Fig. 2(f). In regime I ($\gamma_{dg} + \gamma_{dl} < \gamma_{lg}$), the soil droplet completely spreads, forming a film at the working liquid-gas interface. In II ($\gamma_{dg} + \gamma_{lg} < \gamma_{dl}$), an anti-bubble forms whereby the soil droplet is cloaked by the air, which in turn is surrounded by the invading liquid. In III ($\gamma_{dl} + \gamma_{lg} < \gamma_{dg}$), the droplet disfavors contact with the gas phase and always remains attached to the solid substrate. In IV, V and VI, the three fluid interfaces can form a Neumann triangle, and a capillary bridge is formed between the solid surface and the air interface. The droplet can dewet, as previously illustrated in Fig. 1(b), and we can further distinguish whether we find full (IV) or partial (V) dewetting. In VI, the droplet does not dewet from the surface.

Importantly, we find that dewetting can occur for droplet contact angles significantly lower than 180°. Additionally, as $\theta_{dl}$ increases, the droplet is more likely to dewet across various interfacial tension ratios [Fig. 2(b-e) and S4(b)]. This is expected since higher $\theta_{dl}$ indicates a lower interfacial energy between the working liquid and the surface, compared to the droplet. Dewetting is also preferred for lower $\gamma_{dg}$ and higher $\gamma_{lg}$, which effectively lower $\theta_t$, as they destabilize the submerged droplet and stabilize the liquid lens geometry at the air interface.

We now focus on the sharpness of the boundary between full and non dewetting as predicted by the computational model. When all interfacial tensions are held fixed [arrow in Fig. 2(b)], the transition typically follows a smooth hyperbolic tangent profile when varying the $\theta_{dl}$ [Fig. 2(g)]. In contrast, when $\theta_{dl}$ is held constant and the interfacial tension ratio $\gamma_{dg}/\gamma_{lg}$ varies [arrow in Fig. 2(c)], the transition becomes more abrupt and sharpens further with increasing $\theta_{dl}$ [Fig. 2(h)]. These findings suggest that efficient dewetting can be achieved at relatively low contact angles by tuning interfacial tensions in a ternary system. For instance, abrupt dewetting transitions can be triggered by a slight decrease of $\gamma_{dg}/\gamma_{lg}$.

One important assumption of the model is that of perfectly smooth and homogeneous surfaces. However, experimentally most surfaces exhibit some roughness and chemical heterogeneities. As a result, we typically observe contact line pinning which leads to residues remaining on the surface upon dewetting and a slight deviation in the percentage of removal between experiments and simulations [Fig. 2(a) and SM (8)]. Partial dewetting is more prominent in the experiments than in the simulations.

Another important aspect of surface cleaning is the common use of water-based formulations to enhance outcomes in binary systems. For comparison, we conducted capillary lifting experiments of tetradecane droplets from a PTFE substrate using the low-concentration glycol ether formulation [Fig. 3(a)]. The interfacial tensions $\gamma_{lg}$ and $\gamma_{dl}$ were experimentally measured via the pendent drop [Fig. S6(a) of SM (9)], which is commonly applied to binary fluid systems.

*Contact author: kislon.voitchovsky@durham.ac.uk

†Contact author: halim.kusumaatmaja@ed.ac.uk

However, unlike for the pure liquid systems, when these interfacial tensions are input into the model, the predicted outcomes diverge significantly from experimental results [Fig. 3(a), 1–2]. The associated dewetting prediction is the outlier data point highlighted in Fig. 2(a), confirming that simple pairwise interfacial tension-based predictions for pure liquid no longer hold for complex ternary systems. The insufficiency of pairwise tensions suggests that additional mechanisms are at play.

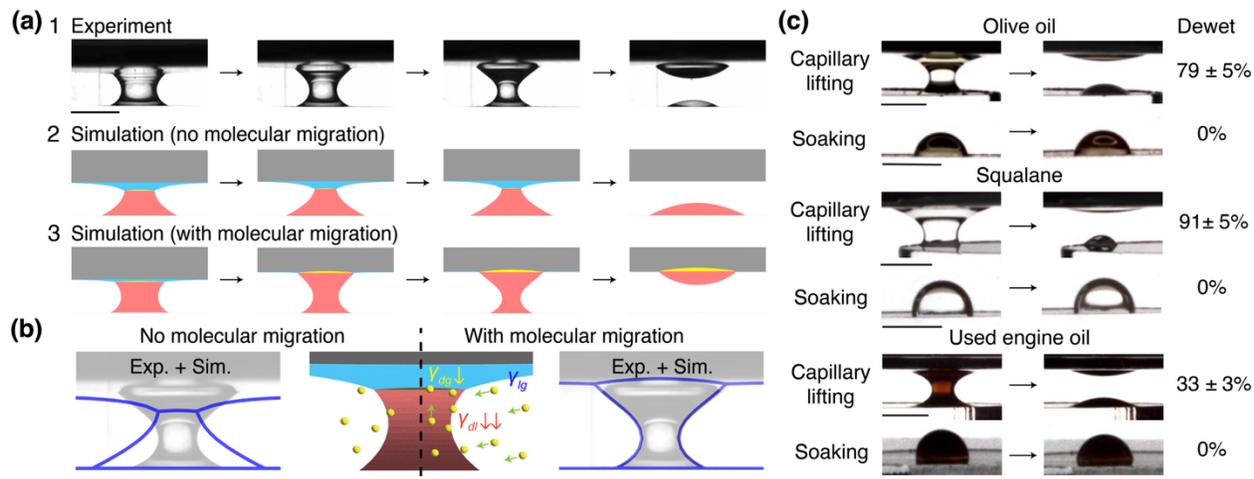

FIG 3. Capillary lifting mechanism in complex systems. (a) Dewetting using a low-concentration formulation as the working liquid. Simulations using binary interfacial tensions fail to capture the experimental profile. In contrast, incorporating the impact of molecular migration between interfaces yields good agreement with experiments (see text). (b) Experimental snapshots overlaid with simulation profiles (blue lines), shown with (right) and without (left) molecular migration. A schematic illustrates the mechanism. (c) Comparison of cleaning effectiveness between capillary lifting and traditional soaking for real-world soil droplets. All scale bars represent 3 mm.

To answer this discrepancy, we need to consider molecular migration across the ternary system. When a droplet forms a capillary bridge between the solid and the air interface, the simultaneous increase in liquid-droplet interfacial area and exposure to the hydrophobic air phase can stimulate amphiphile redistribution within a composite solution, here the formulation. This results in different molecular concentrations at the different interfaces. For example, it has been shown that oil vapor in the air phase can promote surfactant adsorption at water–air interfaces through cooperative interactions with oil molecules that thermodynamically stabilize the adsorption layers [24]. Here, the mechanisms and dynamics of surfactant migration are not known a priori. We start from the so-called Hansen solubility parameters [25] which quantify the cohesive energy of a substance and help predict its solubility in another material. We find that the average Hansen distance of the glycol ethers (in the formulation) and the soil oils (tetradecane or decane) is ~2.6 times smaller than the distance to the water phase [SM (10)]. This indicates a strong preference of the glycol ether for the hydrophobic interfaces, therefore influencing the resulting interfacial tensions in the ternary system. The impact of this migration on the resulting interfacial tensions can be characterized with the Neumann triangle method [26], by depositing an oil droplet on the liquid-air interface to mimic ternary dewetting conditions [Fig. S6(b) of SM (9)]. The results demonstrate a shift in the interfacial tensions from $\gamma_{dg}/\gamma_{lg} = 0.94$ and $\gamma_{dl}/\gamma_{lg} = 0.64$ in the binary system, to $\gamma_{dg}/\gamma_{lg} = 0.84$ and $\gamma_{dl}/\gamma_{lg} = 0.24$ in the ternary system where liquids are exposed to air. This change confirms the importance of glycol ether migration despite their relatively low concentration (1.35 wt%). Incorporating the corrected interfacial tensions into our model yields a good agreement between experimental and computational results [Fig. 3(a) and 3(b)].

Having demonstrated that our framework captures the dewetting behavior in both pure liquids and formulations, we aim to move towards real-world cleaning situations through the capillary lifting mechanism. We do this by comparing the dewetting of common soils by capillary lifting and using the traditional soaking method with and without added detergent. For soils, we use olive oil (food residues), squalane (pharmaceutical, personal care and lubricants) and used engine oil (industrial soils with combustion residues). For the solid we use PMMA, a common, medium surface energy substrate, and pure water as the working liquid. For all soils, we can observe removal by capillary lifting, with dewetting efficiencies of 79 ± 5% for olive oil, 91 ± 5% for squalane, and 33 ± 3% for used engine oil. In contrast, when these soils are deposited on substrates in a water bath to simulate traditional soaking, they remain permanently adhered to the substrate [Fig. 3(c)], despite some changes in the contact angles. The lower efficiency for used engine oil is likely due to the presence of solid residues, but still much better than for soaking. Adding high concentrations of surfactant to the soaking bath eventually removes part of the soils but over a timescale of ~1 hour [SM (11)].

*Contact author: kislon.voitchovsky@durham.ac.uk

†Contact author: halim.kusumaatmaja@ed.ac.uk

In summary, we have investigated capillary lifting mechanism for droplet dewetting in ternary system using a combination of experiments and computational modelling. Our results show that dewetting is favored when $\theta_{dl} > \theta_t$. This corresponds to increasing $\theta_{dl}$ and hence decreasing $\gamma_{dg}/\gamma_{lg}$, offering an alternative to the use of surfactants to decrease the interfacial tension of the working liquid where air is available. For formulated fluids that often contain polymers and surfactants, we also show that molecular migration can significantly change the droplet dewetting outcome. Finally, we demonstrate the ability of capillary lifting to achieve significantly higher cleaning efficiency than soaking for real-world soils. This study advances the understanding of ternary dewetting mechanisms and highlights the potential of capillary lifting as an effective cleaning mechanism for situations with minimal energy and material input. Future studies will investigate the exploitation of related capillary-based phenomena to displace materials from solid surfaces, including solids, films, gels, and other viscoelastic materials.

*Acknowledgments*— KS thanks Abhinav Naga for advice on experimental illumination. This work is supported by EPSRC Centre for Doctoral Training in Soft Matter for Formulation and Industrial Innovation (SOFI2, EP/S023631/1). HK acknowledges funding from UKRI Engineering and Physical Sciences Research Council (EP/V034154/2) and Leverhulme Trust (Research Project RPG-2022-140). KV acknowledges funding from UKRI Engineering and Physical Sciences Research Council (EP/S028234/1).

*Contact author: kislon.voitchovsky@durham.ac.uk

†Contact author: halim.kusumaatmaja@ed.ac.uk

*Contact author: kislon.voitchovsky@durham.ac.uk

†Contact author: halim.kusumaatmaja@ed.ac.uk

**Supplemental Material**

**Droplet Removal by Capillary Lifting**

Ke Sun,[1] Jack R. Panter,[2] Alvin C. M. Shek,[1] Yonas Gizaw,[3] Kislon Voïtchovsky,[1*] and Halim Kusumaatmaja[4†]

[1]Department of Physics, Durham University, Durham, DH1 3LE, United Kingdom
[2]School of Engineering, Mathematics and Physics, University of East Anglia, Norwich, NR4 7TJ, United Kingdom
[3]GreenChemSolution, West Chester, OH 45069, United States of America
[4]Institute for Multiscale Thermofluids, School of Engineering, The University of Edinburgh, Edinburgh, EH9 3FB, United Kingdom

*Contact author: kislon.voitchovsky@durham.ac.uk
†Contact author: halim.kusumaatmaja@ed.ac.uk

**Content**

1. Experimental setup and materials
2. Contact angle measurement of tetradecane and water in air on polymethyl methacrylate (PMMA)
3. Impact of gravitational forces
4. Computational method
5. Static contact angles of substrates and dewetting behavior
6. Table of the interfacial tensions
7. Impact of curved air interfaces for the regime diagram
8. Experimental dewetting outcome compared with simulation prediction
9. Interfacial measurements in binary and ternary system
10. Hansen distance calculation
11. Soaking experiments with added detergent

## 1. Experimental setup and materials

**Substrate and Device Preparation**

Before each experiment, substrates and the dewetting device are cleaned sequentially with deionized water, isopropanol, and deionized water again, followed by drying with nitrogen gas. The fluid device features a central stage for substrate attachment, surrounded by four symmetric holes around it to ensure uniform fluid distribution [Fig. S1]. A 10 $\mu$L droplet is deposited on the substrate, and the working fluid is injected at a rate of 0.02 mL/min using a syringe pump (Pump 11 Pico Plus Elite, Havard Apparatus, Massachusetts, US), corresponding to a liquid rise velocity of 5.33 ×$10^{-7}$ m/s. The device base measures 25 ×25mm and is capped with either a borosilicate (CM Scientific Ltd., Silsden, UK) or acrylic cubic square tube for visualization.

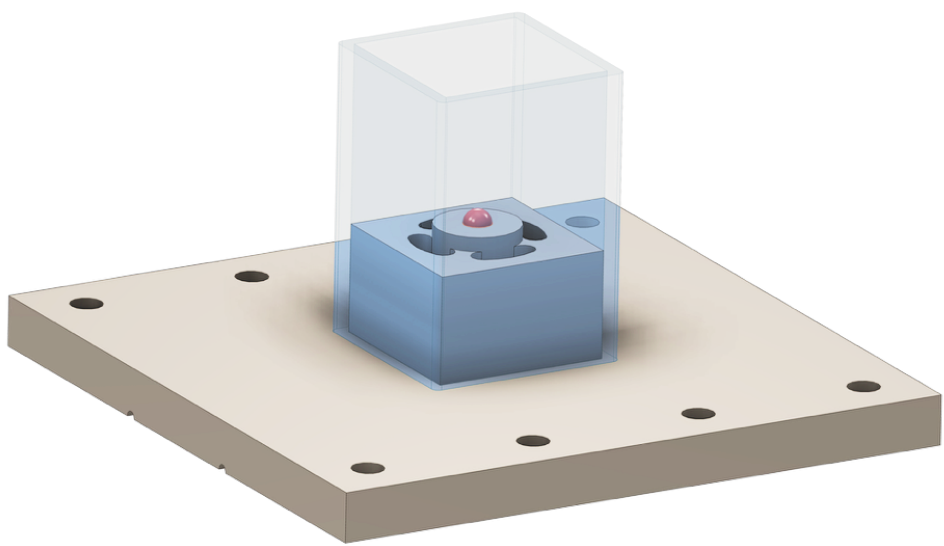

Fig S1. The droplet dewetting device. The central substrate (e.g., stainless steel, as shown) holds the soil droplet, while the working fluid is injected from below through four symmetric holes, ensuring uniform distribution.

**Imaging system**

The dewetting process was captured using: A color digital camera (GV-79L0WP-C-HQ, IDS Imaging, Obersulm, Germany) with a 35 mm HP series lens (Edmund Optics Ltd., York, UK); A mono digital camera (UI-3880CP-M-GL Rev. 2, IDS Imaging, Obersulm, Germany) with a 6.5× zoom lens (MVL6X12Z, Thorlabs, New Jersey, US) and a 0.5× magnifying lens (MVL6X05L, Thorlabs, New Jersey, US). Lighting is provided by either an LED (RALENO, Amazon, UK) or a torch (ICEFIRE, Amazon, UK), coupled with a diffusive screen to ensure uniform illumination.

**Substrates and Liquids**

Substrates include stainless steel (SPI Supplies, West Chester, US), glass (LaboQuip, London, UK), PMMA (Sheet Plastics, Leicester, UK), PC (Sheet Plastics, Leicester, UK), and PTFE (Merck, Sigma-Aldrich, Gillingham, UK). Liquids used are deionized water (18.2 mW, Merck-Millipore, Hertfordshire, UK), tetradecane (Merck, Sigma-Aldrich, Gillingham, UK), decane (Merck, Sigma-Aldrich, Gillingham, UK), silicone oil (10 cSt, Merck, Sigma-Aldrich, Gillingham, UK), olive oil (Solesta, Aldi, Essen, Germany), squalane (Merck, Sigma-Aldrich, Gillingham, UK), and isopropanol (99.5% for HPLC, Thermo Scientific™, Cambridge, UK). For visualization

Fig. 3 of the main text, tetradecane droplets are dyed with Sudan IV (Merck, Sigma-Aldrich, Gillingham, UK) at a concentration of 1mg in 6mL, while water droplets are dyed with fluorescein sodium salt (Merck, Sigma-Aldrich, Gillingham, UK) at a concentration of 100 ppm. The formulation is composed of tripropylene Glycol Methyl Ether 0.45 wt%, diethylene glycol hexyl ether 0.45 wt%, ethylene glycol hexyl ether 0.45 wt%, and deionized water.

## 2. Contact angle measurement of tetradecane and water in air on polymethyl methacrylate (PMMA)

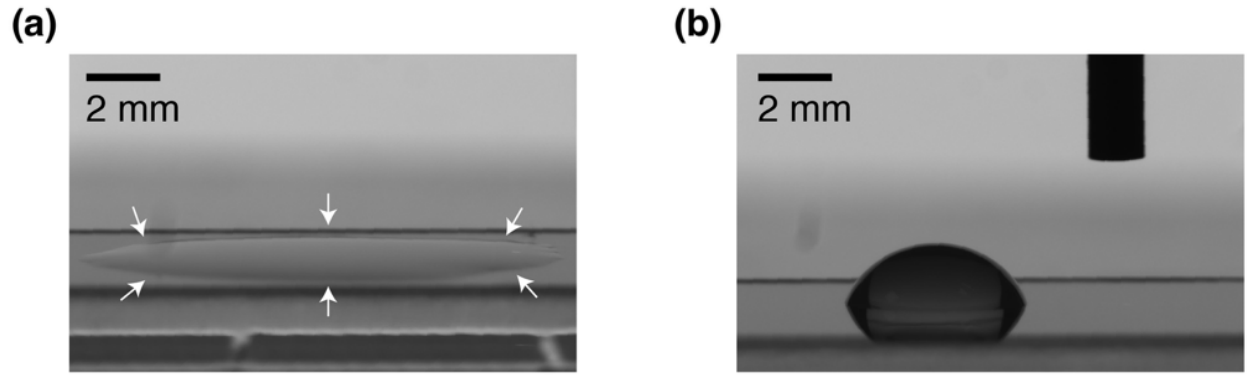

Fig S2. Contact angle measurements of 10 $\mu$L droplets on polymethyl methacrylate (PMMA) substrates: (a) tetradecane and (b) water. White arrows in (a) are added to aid visualization.

As shown in Fig. S2(a), tetradecane exhibits near-complete wetting on PMMA, corresponding to a contact angle of approximately 0°. In contrast, the contact angle of a water droplet on PMMA in air is measured to be 71.22 ± 3.88° [Fig. S2 (b)] by the LBADSA method [1].

## 3. Impact of gravitational forces

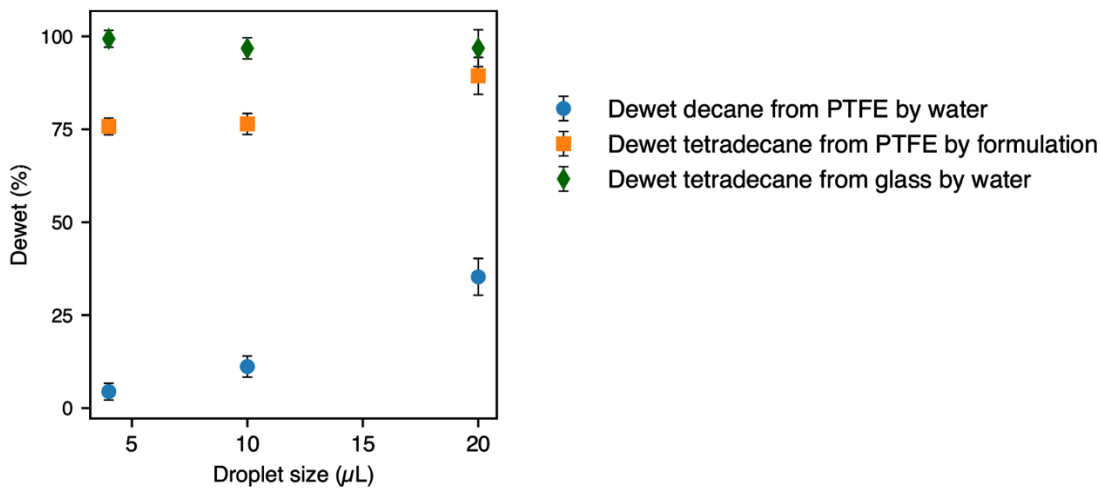

Fig S3. Influence of droplet size on dewetting behavior across different systems. The increased dewetting for larger droplets indicate that gravity cannot be neglected.

The capillary length of all the systems in this work ranges from 1.87 to 4.94 mm and the droplet volume is 10 $\mu$L, making the capillary length comparable in magnitude to the droplet size. To assess the influence of gravitational forces, we experimentally varied the droplet size across different systems. As shown in Fig. S3, increasing the droplet size leads to a higher dewetting percentage. This trend is expected, as the droplets in these systems are less dense than the working liquid, allowing gravity to enhance dewetting for larger droplets. In systems exhibiting near complete dewetting (green, Fig. S3), the effect of droplet size becomes negligible. This is because the system is already in a fully dewetting regime therefore the contribution of gravity is minimized. These observations highlight the necessity of considering gravitational effects, which have been incorporated into our computational model.

## 4. Computational method

We employed Surface Evolver [2] to model the dewetting behavior of a droplet lifted from a solid substrate by an invading fluid under quasi-static conditions. In Surface Evolver, each surface or interface is discretized into triangular facets, and interfacial tensions are assigned based on experimental measurements obtained via techniques such as the pendent drop method or Neumann triangle analysis.

For the solid substrate and side walls, we ensured the chosen interfacial tensions obey the Young's equation. For the droplet on the solid substrate, the relation is given by: $\gamma_{sd} = \gamma_{sl} - \gamma_{dl} \times \cos\theta_{dl}$, where $\gamma_{sd}$, $\gamma_{sl}$, $\gamma_{dl}$ represent the interfacial tensions (or energies) for the solid-droplet, solid-liquid (working fluid), and droplet-liquid (working fluid) interfaces, respectively. $\theta_{dl}$ is the receding contact angle of the droplet on the substrate in the presence of the working fluid. For the working liquid on the side walls, the contact angle is set to $\theta_{lw} = 90°$, such that $\gamma_{lw} = \gamma_{wg} - \gamma_{lg} \times \cos\theta_{lw}$ leads to $\gamma_{lw} = \gamma_{wg}$, where $\gamma_{lw}$ and $\gamma_{wg}$ refer to the interfacial tensions (or energies) for the liquid-wall and wall-gas interfaces, respectively.

In the model, measured interfacial tensions $\gamma_{dg}$ and $\gamma_{dl}$ are normalized by $\gamma_{lg}$ to before being input into the simulation, with $\gamma_{sl}/\gamma_{lg}$ and $\gamma_{wg}/\gamma_{lg}$ set to 1. Densities are normalized by the density of the working liquid. To match the experimental conditions, the gravitational strength $G$ is set by matching the Bond number ($Bo$) of the experiment. The droplet volume remains constant, consistent with experimental conditions. The initial volume of the invading fluid is set to establish a stable liquid bridge between the air and the substrate. The fluid is then added incrementally, followed by energy minimization at each step. The equilibrium state from each step serves as the initial condition for the next, simulating the continuous invasion process under quasi-static conditions.

## 5. Static contact angles of substrates and dewetting behavior

This study employed commercially available substrates to investigate their dewetting behavior, contributing to potential real-life applications. The selected substrates range from hydrophobic (e.g., PTFE) to hydrophilic (e.g., glass and stainless steel), covering a broad spectrum of surface wettability. The static contact angles of hydrophobic droplets on different substrates surrounded by aqueous working fluid illustrate this wettability variation [Fig. S4(a)]. Higher static contact angles indicate weaker droplet adhesion, which leads to increased dewetting. This trend is evident across different systems [Fig. S4 (b)], supporting the choice of surfaces used in this study.

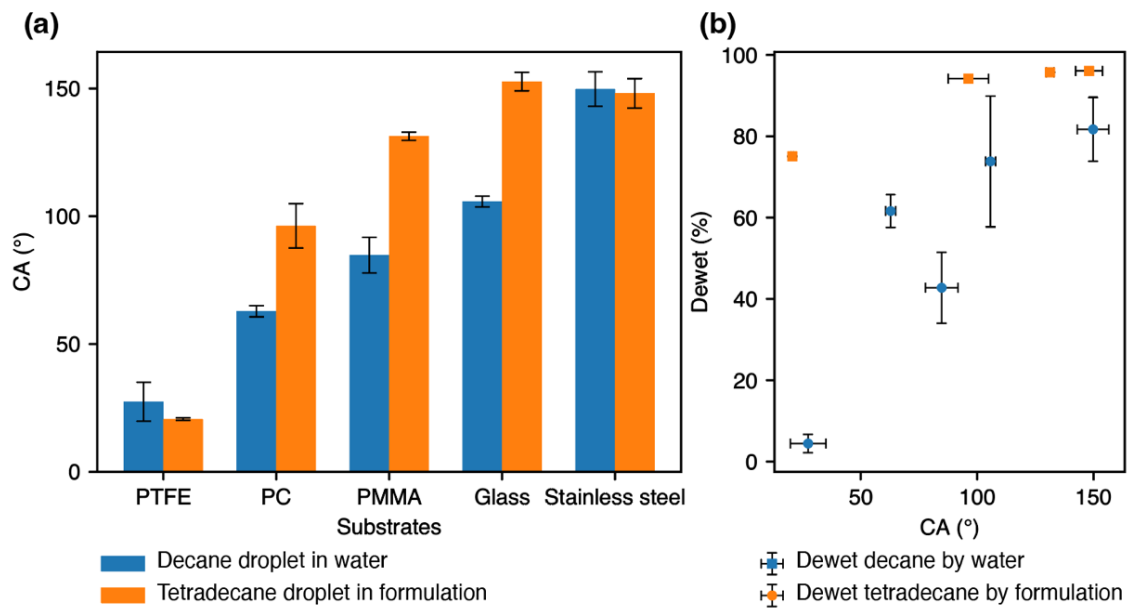

Fig S4. (a) Static contact angles of droplets on various substrates surrounded by the working fluid. (b) The dewetting behavior of droplets as a function of static contact angle.

## 6. Table of the interfacial tensions

The interfacial tensions of liquid-gas and liquid-liquid systems were measured using the pendent drop method. A 22G metal dispense needle (Adhesive Dispensing Ltd., Milton Keynes, UK) was employed, and visualization was carried out as described in the experimental section. Images were processed using the pendent drop plugin [3] in ImageJ freeware [4]. The measured surface and interfacial tensions are provided in Table I.

TABLE I. Measured interfacial tensions.

| Phase I | Phase II | Interfacial tension (mN/m) | Standard deviation (mN/m) |
|---|---|---|---|
| Air | Tetradecane | 26.56 | 0.85 |
| | Decane | 23.62 | 0.32 |
| | Formulation | 28.30 | 0.30 |
| | Olive oil | 32.73 | 0.75 |
| | Squalane | 26.14 | 0.58 |
| | Water (fluorescein dyed) | 73.36 | 0.91 |
| | Tetradecane (Sudan IV dyed) | 24.52 | 0.24 |
| | Silicone oil | 19.48 | 0.26 |
| Water | Tetradecane | 45.20 | 1.37 |
| | Decane | 36.67 | 0.39 |
| | Tetradecane (Sudan IV dyed) | 42.33 | 0.17 |
| | Olive oil | 35.06 | 1.41 |
| | Squalane | 46.11 | 1.54 |
| | Silicone oil | 35.22 | 0.57 |
| Tetradecane | Water (fluorescein dyed) | 36.82 | 2.76 |
| | Formulation | 18.15 | 0.38 |

## 7. Impact of curved air interfaces for the regime diagram

The theoretical line shown in the regime diagram in Fig. 2(a) of the main text corresponds to $\cos\theta_{dl} = \cos\theta_t$, where we have approximated $\cos\theta_t \approx (\gamma_{lg} - \gamma_{dg})/\gamma_{dl}$. This simple geometrical model assumes the liquid-air and droplet-air interfaces to be flat, while in practice these interfaces can be curved. Consequently, deviations can be observed close to the boundary line between dewetting and non-dewetting regimes. To illustrate this, Fig. S5 examines the transition region in detail for fixed interfacial systems with varying $\theta_{dl}$. The red dashed lines indicate contact angles satisfying $\cos\theta_{dl} = \cos\theta_t$, while the black dash lines correspond to 50% dewetting. Small deviations of <10° are typical, which explain the good agreements between experiments, simulations, and the simple model in Fig. 2 of the main text.

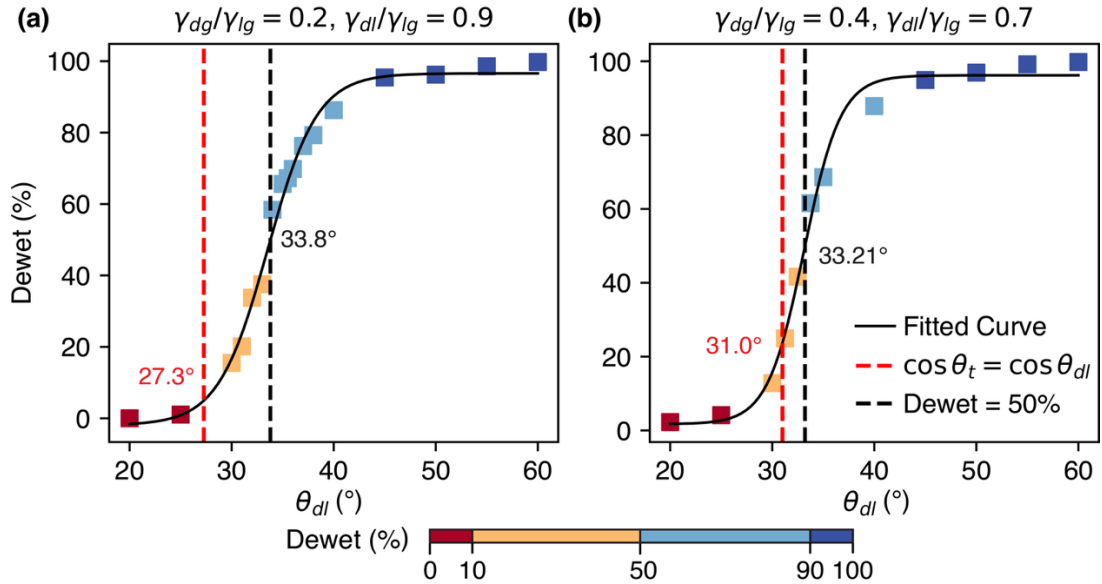

Fig S5. Dewetting and non-dewetting boundaries for fixed interfacial systems: (a) $\gamma_{dg}/\gamma_{lg} = 0.2$, $\gamma_{dl}/\gamma_{lg} = 0.9$; and (b) $\gamma_{dg}/\gamma_{lg}=0.4$; $\gamma_{dl}/\gamma_{lg}=0.7$. The bottom effective contact angle $\theta_{dl}$ is varied, and the data points are fitted with a hyperbolic tangent function. The theoretical boundary $\cos\theta_{dl} = \cos\theta_t$ is shown as a red dashed line, while the simulation predicted 50% dewetting threshold is indicated by a black dashed line.

## 8. Experimental dewetting outcome compared with simulation prediction

TABLE II. Measured dewetting results and simulation prediction.

| Droplet | Working liquid | Substrate | Experiment dewet ± standard deviation (%) | Simulation predicted dewet (%) |
|---|---|---|---|---|
| Decane | Water | PTFE | 4.45 ± 2.25 | 0 |
| Decane | Water | Stainless steel | 81.69 ± 7.83 | 100 |
| Decane | Water | PC | 61.59 ± 4.05 | 100 |
| Decane | Water | PMMA | 86.14 ± 3.63 | 100 |
| Decane | Water | Glass | 73.82 ± 15.04 | 100 |
| Tetradecane | Formulation | PTFE | 78.87 ± 3.11 | 100 |
| Tetradecane | Formulation | Stainless steel | 96.06 ± 0.47 | 100 |
| Tetradecane | Formulation | PC | 94.16 ± 7.83 | 100 |
| Tetradecane | Formulation | PMMA | 95.72 ± 0.44 | 100 |
| Tetradecane | Water | PMMA | 92.11 ± 0.42 | 100 |
| Tetradecane | Water | Glass | 96.77 ± 1.46 | 100 |
| Tetradecane (Sudan IV dyed) | Water | PMMA | 93.35 ± 1.40 | 100 |
| Silicone oil | Water | Glass | 88.30 ± 5.79 | 100 |
| Squalane | Water | PMMA | 79.14 ± 5.24 | 100 |
| Olive oil | Water | PMMA | 90.98 ± 4.77 | 100 |
| Water (fluorescein dyed) | Tetradecane | PMMA | 0 ± 0 | 0 |
| Water (fluorescein dyed) | Tetradecane | Glass | 0 ± 0 | 0 |

Dewetting experiments were performed across a range of fluids and substrates, with each experiment repeated at least 3 times to obtain standard deviations. While most experiments are consistent with simulations, slight deviations are observed (Table II). This discrepancy arises from the fact that simulations exhibit nearly binary behavior, e.g. either complete dewetting or non-dewetting, with a sharp transition in between (see Fig. 2 of the main text). In contrast, experimental outcomes can be affected by substrate heterogeneity, surface roughness, and dynamic effects during the final pinch-off of the capillary bridge, all of which contribute to deviations from the quasi-static simulation predictions on perfectly smooth and homogenous surfaces.

## 9. Interfacial measurements in binary and ternary system

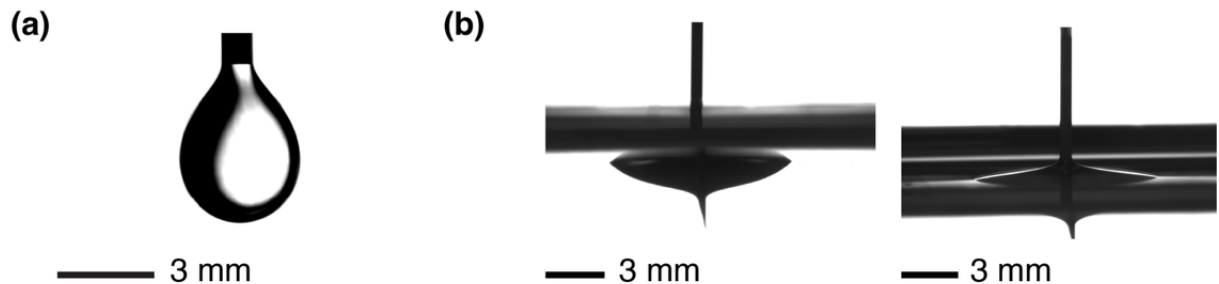

Fig S6. (a) Pendent drop measurement of a tetradecane droplet suspended in the formulated working fluid. (b) Neumann triangle analysis of a droplet of tetradecane floating as a liquid lens at the interface between the formulated working fluid and air.

Interfacial tension measurements were performed using the pendent drop method for binary systems and Neumann triangle analysis for ternary systems [Fig. S6]. In the binary system, a tetradecane droplet is suspended from a needle immersed into the working fluid. The droplet shape is determined by the balance between interfacial tension and gravity and used to extract the interfacial tension. The ternary system consists of the droplet, the working fluid, and the surrounding air. The Neumann triangle method is used to determine the relative interfacial tensions from the contact angles. We get: $\gamma_{dg} : \gamma_{lg} : \gamma_{dl} = 1:1.17:0.279$.

## 10. Hansen distance calculation

To evaluate the solubility of the between components in this multiphase system where glycol ethers formulated in the working fluid, we employed Hansen solubility parameter (HSP) analysis to assess the likelihood of glycol ether migration across the oil–water interface. The interaction distances $R_{a(A-B)}$ of two molecules A and B in Hansen space is calculated using Eq. S1:

$$R_{a(A-B)} = \sqrt{4(\delta_{d,A} - \delta_{d,B})^2 + (\delta_{p,A} - \delta_{p,B})^2 + (\delta_{h,A} - \delta_{h,B})^2} \tag{S1}$$

where $\delta_d$, $\delta_p$, and $\delta_h$ are contributions to the total cohesive energy density from dispersion forces, dipole–dipole interactions, and hydrogen bonding, respectively.

TABLE III. Hansen solubility parameters.

|  | HSP ($[J/cm^3]^{1/2}$) | | |
| --- | --- | --- | --- |
|  | $\delta_d$ | $\delta_p$ | $\delta_h$ |
| Diethylene glycol monoethyl ether (Carbitol ™) | 16.1[a] | 9.2[a] | 12.2[a] |
| Tripropylene glycol methyl ether (DOWANOL™ TPM) | 15.1[b] | 3.5[b] | 11.5[b] |
| Ethylene glycol monohexyl ether (Hexyl CELLOSOLVE™) | 16.2[c] | 9[c] | 5.5[c] |
| Decane | 15.7[d] | 0[d] | 0[d] |
| Tetradecane | 16.2[d] | 0[d] | 0[d] |
| Water | 15.5[d] | 16[d] | 42.3[d] |

[a] from Dow database, Technical Data Sheet, https://www.dow.com/en-us/document-viewer.html?docPath=/content/dam/dcc/documents/110/110-00978-01-carbitol-solvent-tds.pdf
[b] from Dow database, Technical Data Sheet, https://www.dow.com/en-us/document-viewer.html?docPath=/content/dam/dcc/documents/110/110-00619-01-dowanol-tpm-tds.pdf
[c] from Dow database, Technical Data Sheet, https://www.dow.com/en-us/document-viewer.html?docPath=/content/dam/dcc/documents/110/110-00971-01-hexyl-cellosolve-solvent-tds.pdf
[d] taken from Ref. [5].

Using the data from Table III, the calculated Hansen distance $Ra_{(A-B)}$ between the glycol ethers and the oil phase (decane or tetradecane) range from 10.5 to 15.3 $(J/cm^3)^{1/2}$, with an average of 12.66 $(J/cm^3)^{1/2}$. In contrast, the distance to water is significantly larger, ranging from 30.9 to 34.9 $(J/cm^3)^{1/2}$, with an average of 33.03 $(J/cm^3)^{1/2}$. This approximately 2.6 times lower Hansen distance to the oil indicates a much higher solubility of the glycol ethers in the oil phase, suggesting that they preferentially migrate toward the soil droplet.

## 11. Soaking experiments with added detergent

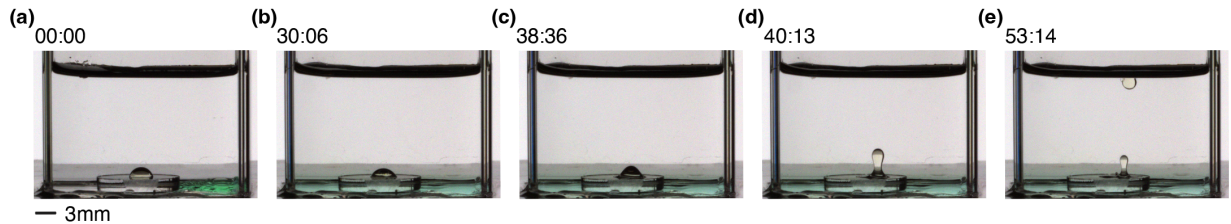

Fig S7: (a) A 0.1 mL solution of 50 w/w% Fairy in water is added to a 10 mL water bath containing a 10 µL olive oil droplet deposited on a PMMA substrate at the bottom. (b) Within the first 30 minutes, the washing-up liquid diffuses around the bottom of the container, lowering the water-droplet interfacial tension and causing the droplet to spread. (c) Then, the washing-up liquid deforms the upper curvature of the droplet, initiating detachment from the central top. This process continues until a significant portion of the droplet is lifted and detached, as shown in (d). After some time, the effect repeats, detaching a second portion of the droplet (e). Timestamps are shown in mm:ss.

The soaking experiment in Fig. 3 of the main text uses pure water as the working fluid. However, in real-world applications, it is common to add washing detergents to the water bath to aid soil removal. To evaluate this scenario, we added 0.1 mL of a 50 w/w% solution of Fairy washing-up liquid to a 10 mL water bath containing a 10 µL olive oil droplet representing the soil, see Fig. S7. We observe that the detergent took approximately 40 minutes to reach the droplet and an additional 10 minutes to remove it, typically in two or more stages. This process is significantly slower than the capillary lifting strategy presented in this work, where around 80% of the droplet is lifted within seconds without surfactant input.